\documentclass[trackchanges,twocolumn]{aastex7}

\usepackage{subfig}
\usepackage{amsmath}

\newcommand{\R}{\textcolor{red}}
\newcommand{\B}{\textcolor{blue}}

\newcommand  *{\diff}   {\mathop{}\!\mathrm{d}}
\renewcommand*{\vec}[1] {\boldsymbol{#1}}
\newcommand  *{\uvec}[1]{\hat{\vec{#1}}}
\newcommand  *{\s}[1]   {\mathsf{#1}}
\newcommand  *{\mat}[1] {\vec{\s{#1}}}
\newcommand  *{\Ups}    {\Upsilon}
\newcommand  *{\The}    {\Theta}
\newcommand  *{\I}      {\mathrm{i}}
\newcommand  *{\Exp}[1] {\mathrm{e}^{\textstyle #1}}
\newcommand  *{\pfrac}[2] {\left(\frac{#1}{#2}\right)}
\newcommand   {\mbhm}{\mathrm{M}_{\mathrm{BH}}}
\newcommand   {\mbh}{$\mathrm{M}_{\mathrm{BH}}$}

\newcommand{\dNprpalt}{de Nicola et al. submitted}

\begin{document}

\title{Eight New Ultramassive Black Hole Masses confirm Best Correlation with Galaxy Core Sizes}

\author{Stefano de Nicola}
\affiliation{Universit{\"a}ts-Sternwarte Muenchen, Scheinerstrasse 1, D-81679, Munich, Germany}
\affiliation{Max-Planck Institute for Extraterrestrial Physics, Giessenbachstrasse 1, D-85748, Garching (Germany)}
\email[show]{denicola@mpe.mpg.de}

\author{Jens Thomas}
\affiliation{Max-Planck Institute for Extraterrestrial Physics, Giessenbachstrasse 1, D-85748, Garching (Germany)}
\affiliation{Universit{\"a}ts-Sternwarte Muenchen, Scheinerstrasse 1, D-81679, Munich, Germany}
\email{jthomas@mpe.mpg.de}

\author{Roberto P. Saglia}
\affiliation{Max-Planck Institute for Extraterrestrial Physics, Giessenbachstrasse 1, D-85748, Garching (Germany)}
\affiliation{Universit{\"a}ts-Sternwarte Muenchen, Scheinerstrasse 1, D-81679, Munich, Germany}
\email{saglia@mpe.mpg.de}

\author{Matthias Kluge}
\affiliation{Max-Planck Institute for Extraterrestrial Physics, Giessenbachstrasse 1, D-85748, Garching (Germany)}
\email{mkluge@mpe.mpg.de}

\author{Jan Snigula}
\affiliation{Max-Planck Institute for Extraterrestrial Physics, Giessenbachstrasse 1, D-85748, Garching (Germany)}
\email{snigula@mpe.mpg.de}

\author{Ralf Bender}
\affiliation{Universit{\"a}ts-Sternwarte Muenchen, Scheinerstrasse 1, D-81679, Munich, Germany}
\affiliation{Max-Planck Institute for Extraterrestrial Physics, Giessenbachstrasse 1, D-85748, Garching (Germany)}
\email{bender@mpe.mpg.de}



\begin{abstract}
We analyse black-hole scaling relations at the high-mass end, focusing in particular on the regime of ultra-massive black holes, $\mathrm{M}_\mathrm{BH} > 10^{10}\,\mathrm{M}_\odot$ (UMBHs). Building on the new dynamical black-hole mass measurements presented in our companion paper for a sample of 16 Brightest Cluster Galaxies (BCGs), including 8 newly identified UMBHs based on triaxial Schwarzschild models, we investigate the implications of these measurements for black-hole scaling relations. This first sample of triaxial black-hole mass determinations increases the number of known UMBHs by a factor of two and dramatically increases the constraints for BH mass scaling relations at the high-mass end. We find that BCGs are outliers in the canonical M$_{\mathrm{BH}} \propto \sigma^{4-5}$ relation, while the size of their depleted cores — the central light-deficient region — offers a tighter correlation with M$_{\mathrm{BH}}$ with scatter 0f 0.22 dex in this mass regime, confirming and extending earlier findings that core size provides a robust proxy at the high-mass end. Our measurements smoothly join the trend already established in previous studies. This also holds for tight correlations between core size and sphere-of-influence radius and core size and core density which, along with the correlation between M$_{\mathrm{BH}}$ and the core size, strongly support the black-hole binary model for the formation of the centers of the most massive galaxies, as the observed core sizes are inconsistent with gravitational-wave recoil predictions and the galaxies display central tangential orbital anisotropy expected from scouring.

\end{abstract}



\section{Introduction}

\label{Sec.introduction}
A currently accepted paradigm is that essentially every massive galaxy hosts a Supermassive Black Hole (SMBH).
Given that SMBHs lie at galaxy centers — thus feeding from the same material as their hosts, although not all this material necessarily loses sufficient angular momentum to reach the galaxy center — there is a close connection between the growth of SMBHs and their host galaxies. Indeed, several galaxy properties have been found to participate in tight scaling relations with the black hole mass M${_\mathrm{BH}}$. These include the galaxy velocity dispersion $\sigma$ \citep{Ferrarese00, Gebhardt00_bhsig, Tremaine02} and the mass (or luminosity) of the spheroidal component, M$_\mathrm{bulge}$ (L$_\mathrm{bulge}$, \citealt{Magorrian98, Marconi03}). On the one hand, these scaling relations encode important information about the physical processes that link SMBH and galaxy evolution together over a large range of galaxy and SMBH masses. On the other hand, since direct dynamical SMBH mass determinations are still rare, these scaling relations are important practical tools to approximate black hole masses in large surveys or in galaxies with unresolved centres. Hence, more than one work has raised the question which relation predicts $\mbhm$ most precisely, with different relations emerging as most effective for different galaxy types and mass regimes (e.g. \citealt{Seigar08, Bogdan18}). While $\sigma$ often performs well over the full mass range \citep{Kormendy13, Rob16, VDB16, Shankar16, dN19, Shankar25} and there are works reporting a steepening of this relation at the high-mass end (e.g. \citealt{Dullo19, Sahu19}), its predictive power for BCGs appears weak in our sample, consistent with earlier indications that BCGs may deviate from the canonical scaling relations at the high-mass end (e.g. \citealt{DallaBonta09, Graham13, Bogdan18}), possibly because of different normalizations for different galaxy types \citep{Zubovas12, King26}. \\
At the high-mass end galaxy mergers are no longer dissipational but predominantly gas-free without star formation and without dissipationally-driven feedback processes. Such mergers in general do not significantly increase $\sigma$ \citep{Boylan06, Naab09, Kormendy13FJ, King15}. 
The most striking observational evidence for gas-poor merging is the fact that massive elliptical galaxies exhibit a light deficit at their center where the surface brightness (SB) profile flattens compared to an inward extrapolation of the Sersic \citep{Sersic63} profile characteristic at large radii. Such galaxies are referred to as a "core-elliptical" \citep{Faber97, Lauer07, Rusli13cores, Dullo14, Kianusch19}. The origin of these cores is explained by in-falling SMBHs of the progenitors that eject stars via gravitational slingshots \citep{Begelman80, Ebisuzaki91, Merritt06, Jens14, Rantala18, Rantala19, Frigo21}.

Lying at the centre of the potential well of the cluster, these most massive galaxies experience a different evolution history, and the nature of the black-hole scaling relations changes as well. The cores give rise to new scaling relations holding for core-ellipticals, governed by the dynamics of black-hole binaries. These relations are local, basically confined to the sphere-of-influence of the central black-hole binary or, later, the merged black hole.
Indeed, the amount of light in the core \citep{Graham04, Kormendy09}, the core size \citep{Ferrarese06, Lauer07, Dullo14} and sphere-of-influence size \citep{Jens16}, and the SB of the core itself \citep{Kianusch19} have been discovered to scale tightly with the SMBH mass.
Instead, the global scaling relations between $\mathrm{M}_\mathrm{BH}$ and $\sigma$ are expected to change at the high-mass end, as dry mergers increase galaxy mass and size more efficiently than velocity dispersion, leading to a steepening of the $\mathrm{M}\mathrm{BH}$--$\sigma$ relation and a reduced sensitivity of $\sigma$ to further black-hole growth \citep{Naab09,Kormendy13FJ,Graham23_virial}.

Just like the classical scaling relations between the black hole mass and the velocity dispersion or the stellar mass allow to estimate \mbh\ for intermediate mass early-type galaxies, the local core correlations allow to estimate \mbh\,for massive galaxies, even with less scatter. While more recent \mbh-$\sigma$ and \mbh-$M_\ast$ relations calibrated specifically for core-S{'e}rsic galaxies predict more massive black holes than the classical relations, core-based correlations remain among the tightest predictors at the high-mass end \citep{Shankar25}. Core sizes, for example, have led to the discovery of the most massive SMBH found so far \citep{Kianusch19} in a galaxy where the classical scaling relations predict a SMBH ten times smaller than observed.

While the physical process behind these core correlations -- dynamics of SMBH binaries in gas-poor mergers -- is expected to be generic at the high-mass end, little is known about SMBHs with masses above $10^{10} \mathrm{M}_\odot$, often called ultra-massive black holes (UMBHs). So far, only 7 UMBHs have been discovered \citep{McConnell12, Jens16, Kianusch19, Bianca23N5419, Nightingale23, dN24, MeloCarneiro25}. Here, we analyse SMBH scaling relations for a sample of 16 BCGs that are candidate hosts of UMBHs together with massive core galaxies from the literature \citep{Rusli13cores, Rob16, Jens16, Kianusch19, Rob24}. This is the first sample of UMBHs with triaxial black-hole mass measurements.
Throughout this paper, we identify the BCG by the cluster name.

\section{Sample and data}
Our base sample consists of 17 BCGs from the large collection of 170 local BCGs of \citep{Matthias20, Matthias21} with
very deep $g'$-band Wendelstein \citep{Hopp10, LangBardl16} data. The galaxies were selected to be candidate hosts of UMBHs due to their large central cores larger than 1 kpc or their high mass, both known to correlate with \mbh\,\citep{Jens16, Rob24}. One of them (A262) has already been published in a pilot study \citep{dN24}. Some of these galaxies have archival high-resolution HST data, for the others we obtained AO-aided photometry from the Large Binocular Telescope Observatory (LBTO). In addition, long-slit spectra along 4 different position angles were collected using the MODS instrument at LBTO. We perform triaxial orbit-based dynamical models to measure the masses of the black holes, stars and dark-matter halos \citep{dN20, Bianca21, Mathias21, Jens22, dN22Nbody, Bianca23Nbody}. We complement this sample of galaxies with new black-hole measurements in the UMBHs regime with existing data of other massive core ellipticals from  \citet{Rusli13cores, Jens16, Kianusch19, Rob24}.\\

For 13 BCGs we estimate the core size r$_\mathrm{c}$ using the break radius r$_\mathrm{b}$ through fitting PSF-convolved\footnote{The fits have been performed using the \texttt{curve\_fit} routine of \texttt{scipy} \citep{Virtanen20_scipy} and following the prescriptions of Sec. 4.4 of \citet{Matthias20}.} Core-Sérsic \citep{Sersic63, Graham03, Trujillo04} profiles to the 1D SB profiles of the galaxies, extracted along the semimajor axes of the isophotes. These have been derived using the software of \citet{Bender87}, which allows for varying ellipticity and PA profiles. The Core-S{\'e}rsic function is defined as follows:

\begin{equation}
\begin{aligned}
&\mathrm{SB}(R_{circ}) = \mathrm{SB}_e
- 2.5\left(\frac{\gamma}{\alpha}\right)\,\log_{10}\!\left[1 + \left(\frac{R_b}{R_{circ}}\right)^{\alpha}\right] \\
&+ 2.5\,(0.868\,n - 0.142)\,
\left\{
\left[
\frac{R_{circ}^{\alpha} + R_b^{\alpha}}{R_e^{\alpha}}
\right]^{\!\frac{1}{\alpha n}}
- 1
\right\},
\end{aligned}
\end{equation}

\noindent where $R_{circ}$ is the the circularized radius\footnote{This is evaluated following \citet{Rob10_FP} as $(a+b)/2$, where $a$ and $b$ are the semi-major and -minor axis lengths of the isophotes. Using $\sqrt{ab}$ yields no significant differences.} of the isophotes, $R_e$ is the effective radius, $R_b$ is the break radius, $\gamma$ is the inner slope and $\alpha$ controls the sharpness of the transition\footnote{For 6 BCGs, \citet{Matthias20} report that fitting a Core-S{\'e}rsic + S{\'e}rsic profile reproduces the SB profile better. We show in \citet[App.~B2]{dN26} that using only one Core-S{\'e}rsic function makes no significant difference to the measured core sizes.}. \\
The high-resolution data (HST or AO-assisted LBT, see \dNprpalt\,for details) allow us to resolve the cores. Furthermore, the availability of both F814W HST images and H-Ks LBT images enables us to assess the impact of dust on our measurements. The excellent agreement between the two datasets indicates that our core size estimates are not affected by the presence of a dusty nuclear disk. \\
For A292 and A1185 we measure r$_\mathrm{c}$ by computing the cusp radius r$_\gamma$, defined as dI/dlogr (r$_\gamma$) = -1/2 \citep{Carollo97}, noting that this criterion alone does not necessarily imply the presence of a depleted core, since any Sérsic profile eventually reaches a logarithmic slope of $-1/2$. Nevertheless, based on the detailed photometric analysis presented in the companion paper, we classify both galaxies as genuine core systems. These values are converted into kpc using the distances of Tab.~\ref{Tab.bh_core} and used for all correlations presented throughout the paper. For A2107 only a single Sersic fit is needed to fit the  SB profile, meaning that the galaxy does not have a core. Finally, since A1185 has a companion galaxy located $\sim$7 arcseconds from the centre, we exclude this object from the scaling relations, although we retain it in the dynamical modelling and show it in all figures as a star.\\
We take $\mathrm{M}_\mathrm{BH}$ from the triaxial dynamical models where we also determine the radius of the sphere of influence, $\mathrm{r}_\mathrm{SOI}$, defined as $\mathrm{M}_\mathrm{BH} = \mathrm{M}_\ast(r \leq r_\mathrm{SOI})$, where M$_\ast$ is the stellar mass computed as described in Sec. 4 of \citet{dN26}. We also calculate a velocity dispersion $\sigma$ as

\begin{equation}
   \sigma = \sum_i \nu_i(R_i) \times w_i(R_i)\,/\,\sum_i w_i(R_i),
    \label{eq.sigma}
\end{equation}

\noindent summed over all radial bins within the instrument field-of-view, with $\nu = \sqrt{v^2 + \sigma^2}$ and $v, \sigma$ the measured line-of-sight velocity and velocity dispersion. We take as weights $w_i(R) = 2\pi R I(R)$, where $R$ is the radius of the i-th bin and $I(R)$ the surface intensity. A detailed description of the photometric and spectroscopic data, the Core-Sersic fits, the kinematic extraction of non-parametric LOSVDs with our new fitting code WINGFIT (\citealt{Kianusch23}, Thomas in prep.) and a description of the dynamical models is provided in a companion paper \citep{dN26}.

\section{Scaling relations} \label{Sec.scaling}

Our novel 8 UMBH detections more than double the number of such systems, allowing for a statistically significant investigation of the classical scaling relations between black holes and host galaxies such as the $\mbhm-\sigma$ and those involving core ellipticals at the very high-mass end, providing insights in the coevolution of these systems and their hosts. We analyze the canonical $\mbhm-\sigma$ correlation before focusing on r$_\mathrm{c}$ and its correlations with \mbh, the sphere-of-influence r$_\mathrm{SOI}$, the stellar mass M$_\ast$ and the core density $\rho_\mathrm{core}$, along with a possible correlation between \mbh\,and the mass deficit inside the core M$_\mathrm{def}$. We finally speculate about the existence of UMBHs in core-less galaxies. \\
All linear regressions are performed using the robust routine from \citet{Cappellari13_lts}, which allows for measurement errors on both variables and intrinsic scatter in the y-variable.
All the data needed to generate the plots presented in the next sections are reported in Tab.~\ref{Tab.bh_core}. For all other galaxies we take the respective data from the literature \citep{MeloCarneiro25, Rob24, dN24, Bianca23N5419, Kianusch19, Jens16, Rusli13cores}.

\subsection{The break-down of $M_{BH}-\sigma$} \label{Ssec.bh_sigma}
\begin{figure}
    \centering
    \includegraphics[width=1.0\linewidth]{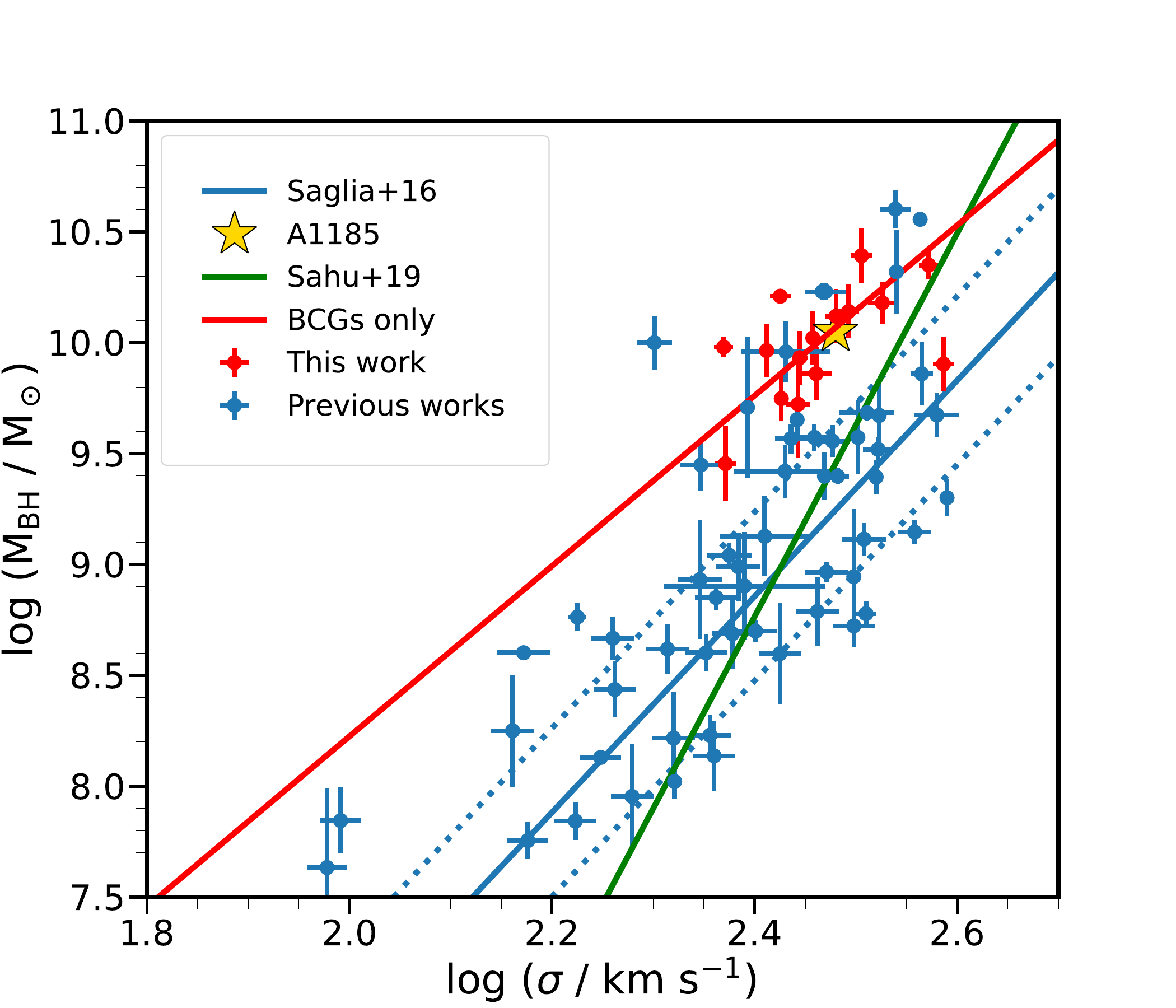}
    \caption{The \mbh-$\sigma$ relation with data from the literature (blue points) with our new measurements (red points). The red line is the relation we find by only considering BCGs. The solid line is the best-fit relation from \citet{Rob16} omitting pseudobulges, while the dotted lines enclose the intrinsic scatter $\varepsilon = 0.380\,\pm 0.038$. Finally, the green line is the steeper relation from \citet{Sahu19} when fitting core-S\'ersic galaxies.}
    \label{Fig.BH_sigma}
\end{figure}

Core ellipticals do not follow the canonical $\mbhm-\sigma$ scaling relation (see, e.g., \citealt{Rob16, Kianusch19}). This breakdown at the high-mass end has been anticipated from quasar energetics and luminosity function arguments (e.g., \citealt{LyndenBell69, Rees84, Netzer03, Lauer07_BHs}) and was quantified observationally by \citet{Dullo19}, who showed that $\mbhm$–$\sigma$ estimates for large-core galaxies can be undermassive by up to a factor of $\sim$40. BCGs, in particular, often show low velocity dispersions (\citealt{Matthias23BCGs}, \dNprpalt): the $\mbhm \propto \sigma^{4-5}$ relation would predict very small black holes with $\mbhm < 10^9$ M$_\odot$ for these galaxies. In Fig.~\ref{Fig.BH_sigma}, we show the dispersions and black-hole masses of our galaxies together with the best-fit relation for less massive galaxies and additional objects from \citet{Rob24} and references therein. We also plot the best-fit line by only considering BCGs as well as a steeper relation found by \citet{Sahu19} for a sample of 35 Core-S{\'e}rsic galaxies. 
The figure confirms that nearly all BCGs from our sample host overmassive BHs relative to the predictions of the canonical \mbh-$\sigma$ relation, and the effect is particularly severe due to the low $\sigma$ values of most BCGs. Low velocity dispersions in massive ETGs fit into the dry merger scenario as discussed in the introduction. Fitting a $\mbhm$–$\sigma$ relation for BCGs only does not improve the situation, yielding an even smaller slope of 3.84 and a Spearman correlation coefficient of 0.49, indicating only a weak correlation. This differs from the steeper relation reported by \citet{Bogdan18}, who found a slope of 9.68 and a Spearman correlation coefficient of 0.58 for their BCG sample.

\subsection{M$_{BH}$-r$_\mathrm{c}$} \label{Ssec.bh_core}
\begin{figure}
    \centering
    \includegraphics[width=1.0\linewidth]{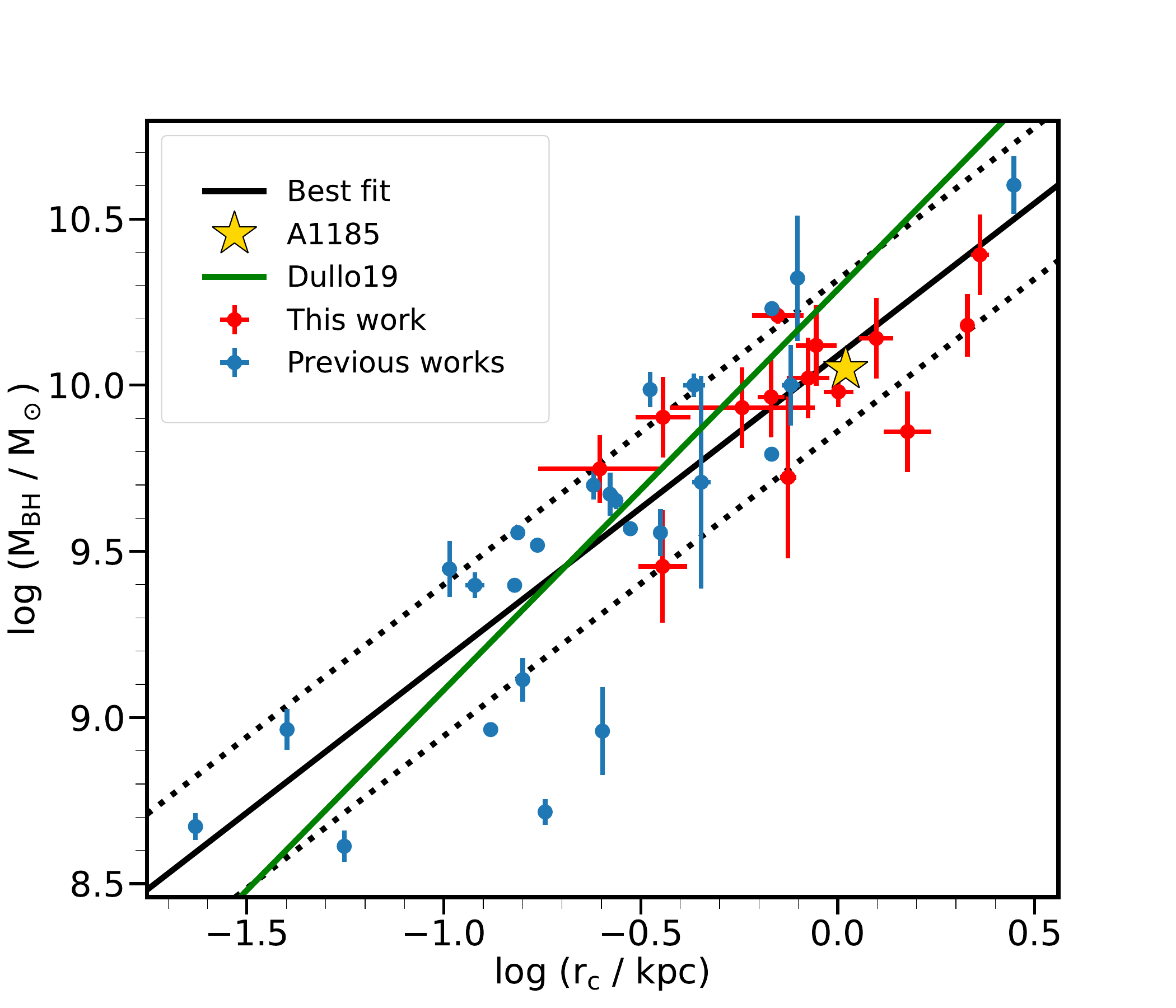}
    \caption{The \mbh-core size relation with data from the literature (blue points) with our new measurements (red circles), along with the relation from \citet{Dullo19} shown as green line. The solid and dotted blue lines are the best-fit line and the intrinsic scatter, respectively.}
    \label{Fig.BH_core}
\end{figure}

An inevitable consequence of dry merging between galaxies with supermassive black holes is core scouring through a combination of dynamical friction and gravitational slingshots \citep{Frigo21}. At the end of this process the remnant galaxy has a diffuse core with a shallow, almost constant surface brightness profile although additional mechanisms involving the infall of satellite galaxies have also been proposed to contribute to the formation of large cores in BCGs \citep{Goerdt10, Bonfini15}.
It can be characterized by a core or break radius that defines the transition from an outer S{\'e}rsic-like surface-brightness distribution to the core. The core typically follows a shallow power-law distribution. Previous studies have shown that massive elliptical galaxies
with depleted stellar cores exhibit an almost linear relationship between the masses $\mathrm{M}_\mathrm{BH}$ of their black holes and the size of the core \citep{deRuiter05, Lauer07_BHs, Jens16, Dullo19}. Fig.~\ref{Fig.BH_core} shows this relation for our new sample, along with the steeper relation found by \citet{Dullo19} for a sample of 11 Core-S{\'e}rsic galaxies. It is clear that the relation extends to the regime of UMBHs. As also shown by \citet{Lauer07}, core size provides an independent and more reliable predictor of $\mathrm{M}\mathrm{BH}$ at the high-mass end than $\sigma$, which saturates in BCGs (Fig.~\ref{Fig.BH_sigma}). Instead, the core size scales as tightly with $\mathrm{M}_\mathrm{BH}$ at the high-mass end, as does the dispersion $\sigma$ for intermediate-mass bulges. \\


We obtain a robust log-linear relation
\begin{equation}
\log \frac{\mathrm{M}_\mathrm{BH}}{\mathrm{M}_\odot} = (0.916 \pm 0.081) \log \frac{\mathrm{r}_\mathrm{c}}{\mathrm{kpc}} + (10.087 \pm 0.053)
\label{eq.rbmbh}
\end{equation}
with an intrinsic scatter of $\varepsilon = 0.222\,\pm 0.035$ and a Spearman's correlation coefficient of 0.89. Nearly all data points fall within the intrinsic scatter of this relation, and the slope is consistent with previous measurements for large-core galaxies \citep{Jens16, Dullo19}. \\
We observe two kinds of outliers in the $\mbhm-\mathrm{r}_\mathrm{c}$ relation. The galaxy A240 is an example of an outlier with an unusually large core. Neither our inspection of the HST images nor the analysis of \citet{Matthias20} reveals evidence for a secondary nucleus. The surface-brightness profile exhibits a slight excess at $\sim$1-2 arcsec, but no corresponding photometric structure is visible in the imaging. We therefore do not interpret this feature as evidence for an additional nucleus. Simulations show that recoil kicks can enlarge cores up to a factor of 3  when the recoil velocity exceeds $\sim$50\% of the escape velocity \citep{Rawlings25}. While this is not the general case for our sample - our black hole masses correlate well with the core size - for galaxies such as A240, this mechanism could play a role. We discuss large BHs found in galaxies with small (or no) cores in Sec.~\ref{Sec.core_less}. \\
We conclude this section by stressing that the black-hole binary model is the only one which is consistent with our observations. Gravitational-wave recoil is expected to produce cores that are systematically larger than those generated by binary scouring alone, whereas the core sizes in our sample are broadly consistent with expectations from the literature SMBH mass–core size relation. A purely dry-merger origin is also disfavoured by the presence of central tangential orbital anisotropy in all BCGs in our sample \citep{dN26}, a dynamical signature naturally predicted by SMBH binary scouring.

\subsubsection{Effect of secondary nuclei}
Possible contamination from secondary nuclei does not affect our conclusions. Although \citet{Matthias20} identified secondary nuclei in 24\% of their BCG sample, nearly half of these cases are consistent with chance superpositions. The galaxies for which \citet{Matthias20} report a double component are A160, A592, A1185 (which we omit), A1775, A2256, A2319 and A2506. \\
The photometry of A160 and A2506 is undisturbed. Likewise, A592 and A2319 show no evidence of disturbed morphology in the LBT images, and omitting them from the scaling relations does not alter our results. The unusually large cores of A1775 and A2256, together with their anisotropy profiles within the core regions, are consistent with past mergers between two core ellipticals. However, the HST photometry of these galaxies shows no evidence for secondary nuclei.

\subsection{$M_{BH} - M_\ast$ and $r_c - M_\ast$} \label{Ssec.mgal}
\begin{figure*}
\centering
\subfloat[\label{Fig.mbh_mbul_G}]{\includegraphics[width=0.4\linewidth]{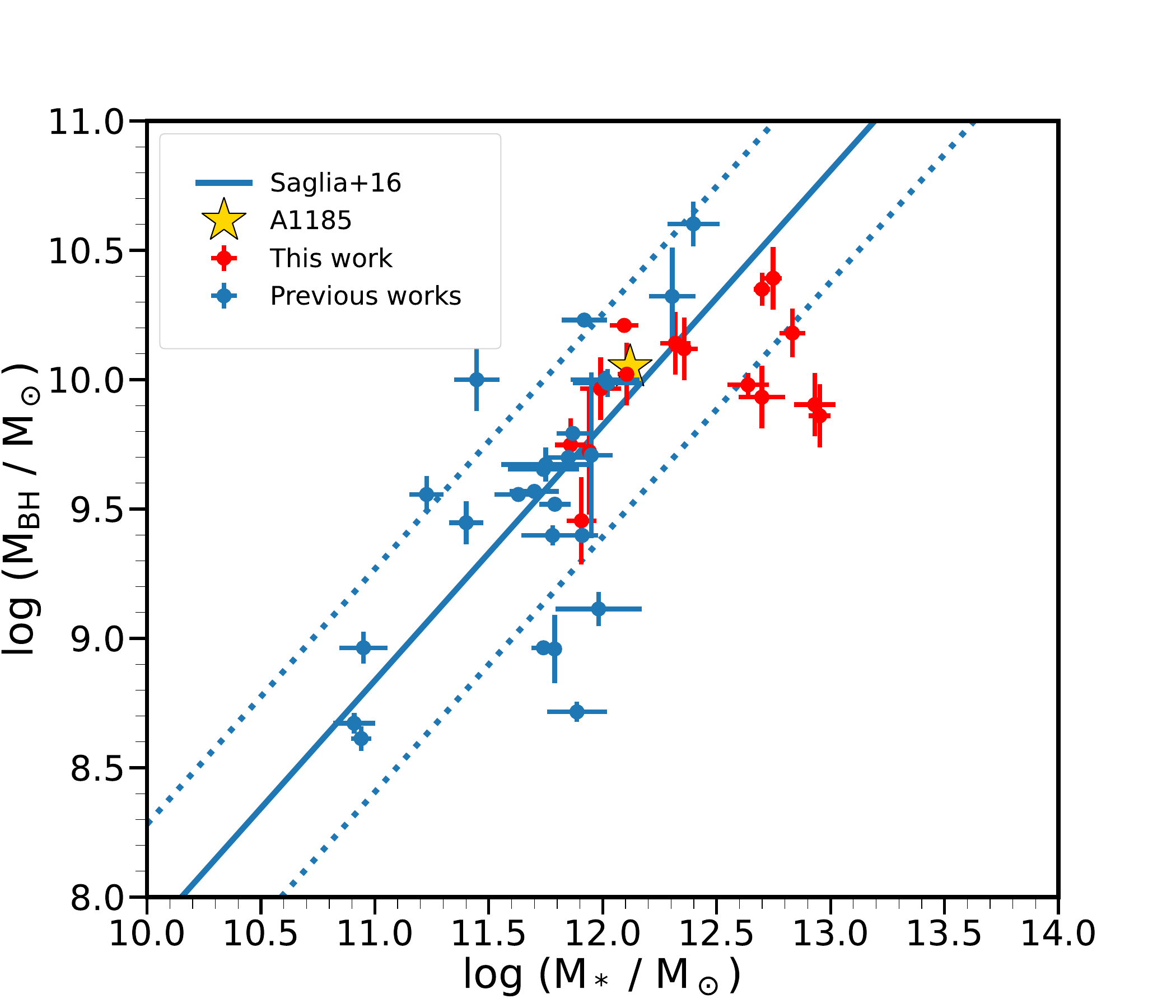}}
\subfloat[\label{Fig.mbh_mbul_U}]{\includegraphics[width=0.4\linewidth]{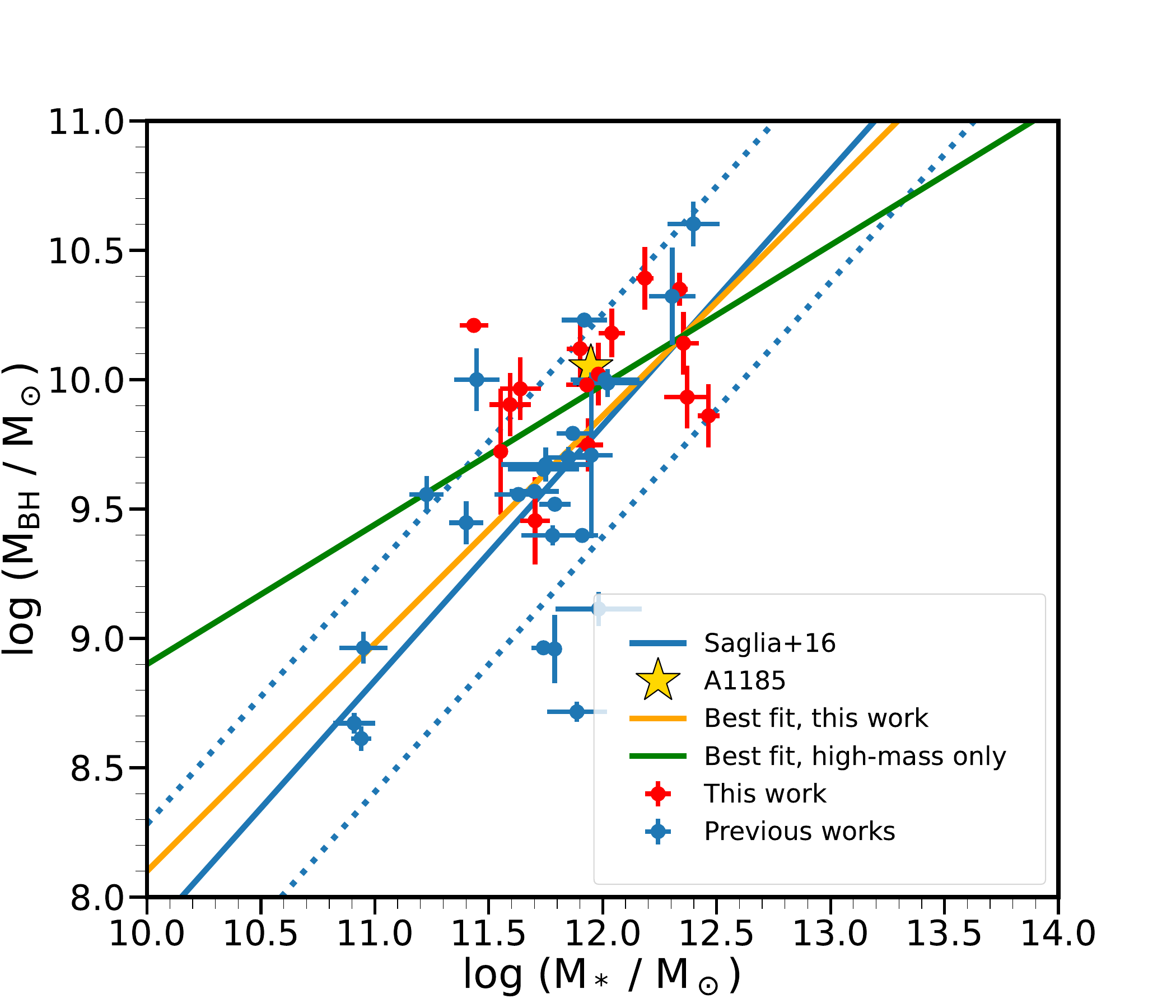}}

\subfloat[\label{Fig.mgal_rb_G}]{\includegraphics[width=0.4\linewidth]{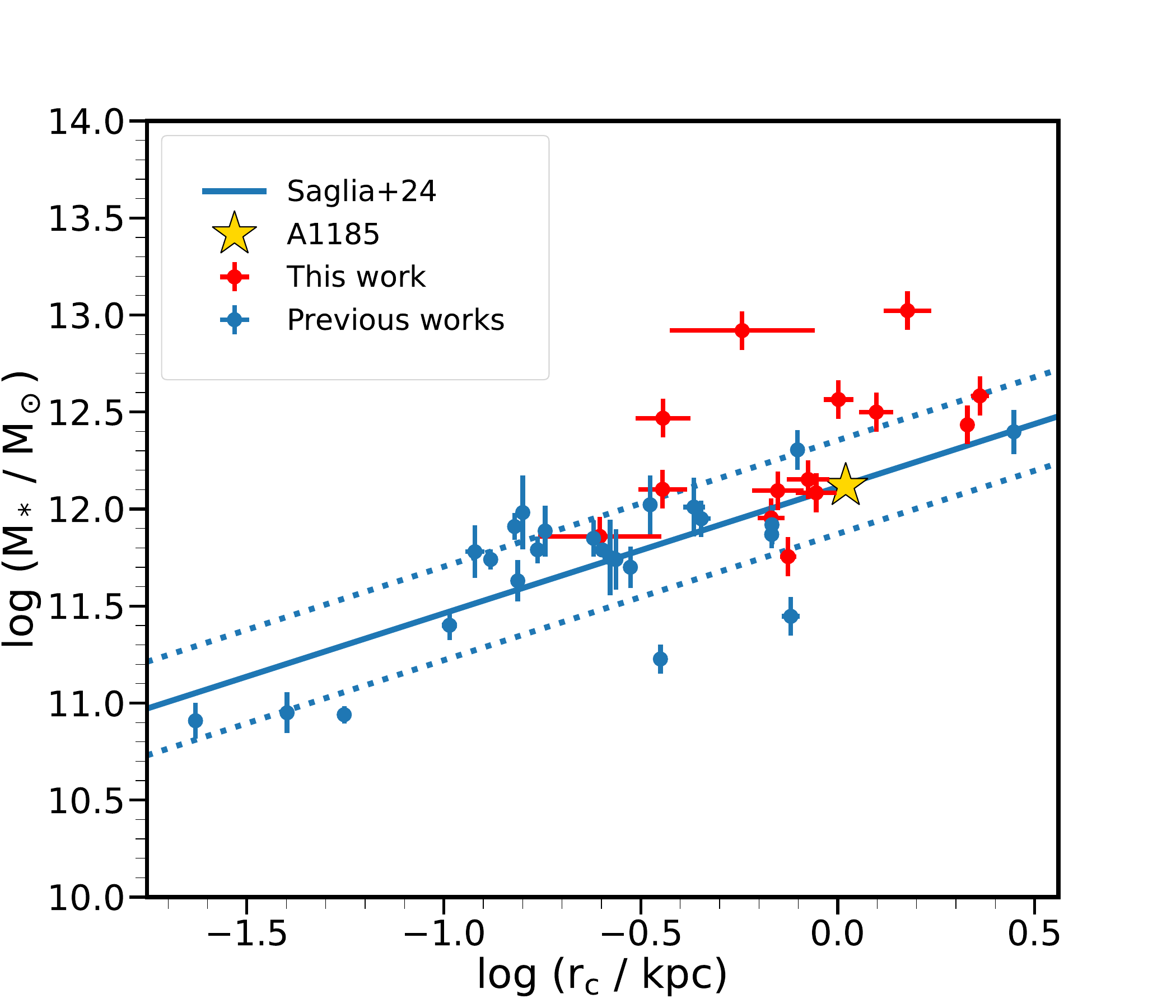}}
\subfloat[\label{Fig.mgal_rb_U}]{\includegraphics[width=0.4\linewidth]{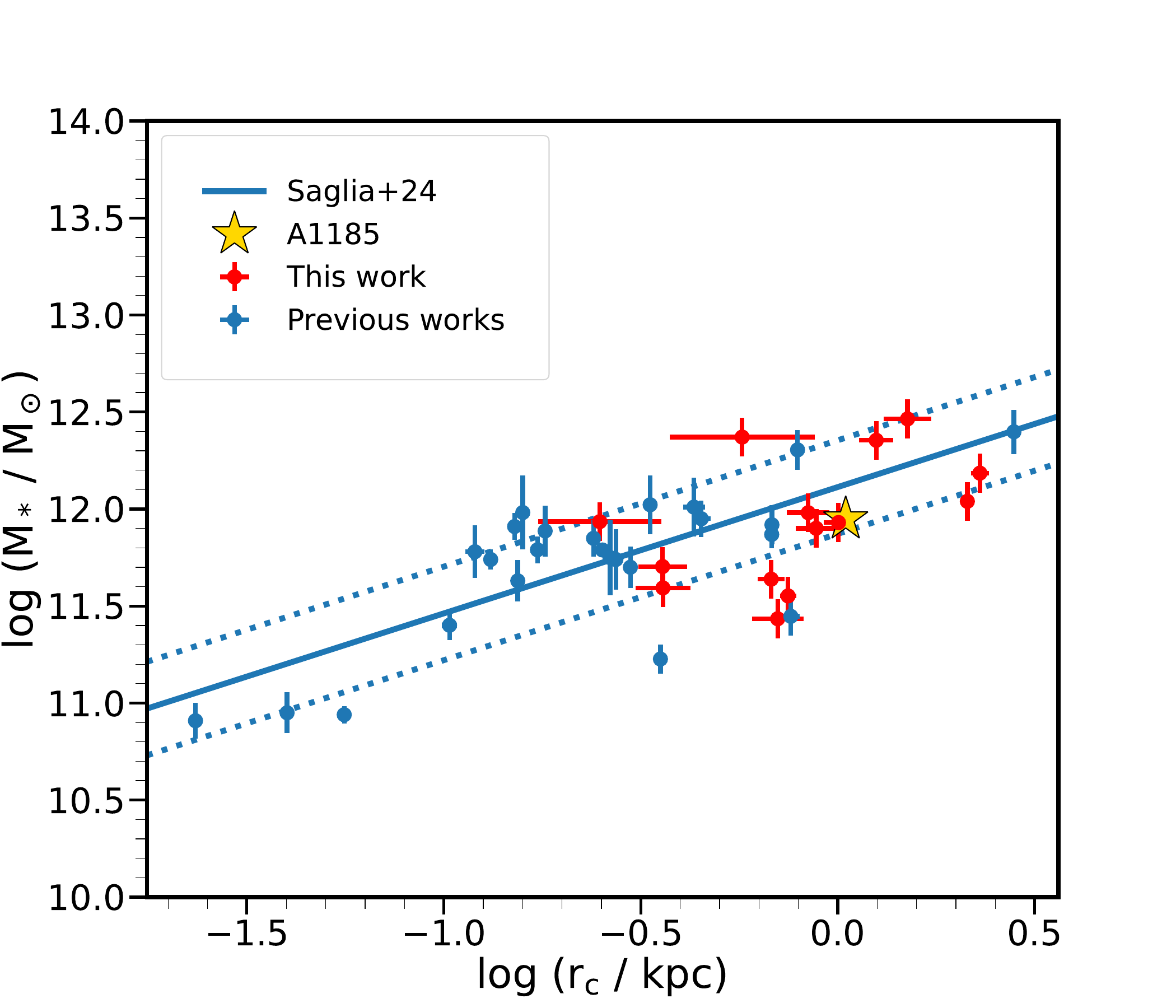}}

    \caption{Same as Fig.~\ref{Fig.BH_sigma} for the $\mathrm{M}_\mathrm{BH} - \mathrm{M}_\ast$ and $\mathrm{M}_\ast - \mathrm{r}_\mathrm{b}$ relations. In Figs.~\ref{Fig.mbh_mbul_G} and~\ref{Fig.mgal_rb_G} the stellar mass has been estimated using the total V-band magnitude of the BCG+ICL, whereas in Figs.~\ref{Fig.mbh_mbul_U} and~\ref{Fig.mgal_rb_U} it comes from the integrated 1D best-fit Sersic profiles, corrected for ICL contamination following \citet{Matthias21}. In both cases the data are taken from Table 4 of \citet{Matthias20}; the best-fit relations and intrinsic scatters (blue lines) come from \citet{Rob16} (top panels) and \citet{Rob24} (bottom panels), respectively. Finally, in Fig.~\ref{Fig.mbh_mbul_U} we show the best-fitting relation for the full galaxy sample (orange) and for the subsample of galaxies with $M_{\rm BH} > 10^{9.5}\,M_\odot$ (green).}
    \label{Fig.mbh_mgal}
\end{figure*}

A classical scaling relation that is obeyed by core ellipticals is that between $\mathrm{M}_\mathrm{BH}$ and the galaxy stellar mass $\mathrm{M}_\ast$ \citep{Rob24}. This tells us that the most massive BHs are found in the most massive galaxies, agreeing well with the picture of coevolution of BH and host galaxies \citep{Kormendy13}. \citet{Lauer07_BHs} have shown that BCGs follow scaling relations as other giant ellipticals, justifying the use of these relations to predict $\mathrm{M}_\mathrm{BH}$ in BCGs. Here, we evaluate $\mathrm{M}_\ast$ in two ways:
\begin{itemize}
    \item We take the total integrated magnitude $\mathrm{M}_V$ of \citet{Matthias20} and turn this into a mass estimate using our dynamically estimated\footnote{The dynamical mass-to-light ratios quoted here refer only to the stellar component. Our Poisson solver explicitly includes a separate dark matter potential ($\rho_{\rm DM}$), while the scaling factor $\Gamma$ is used to normalize the stellar mass distribution independently of the dark matter contribution.} mass-to-light ratios (see \citealt{dN26});
    \item We evaluate the total luminosity by integrating the innermost Sersic component reported by \citet{Matthias20} to fit the SB profile up to infinity. If only one component is found, we correct for the ICL contamination following \citet{Matthias21}.
\end{itemize}
The exact values are reported in Tab.~\ref{Tab.bh_core}. In the top panels of Fig.~\ref{Fig.mbh_mgal} we show the correlation of \citet{Rob16} with the points from \citet{Rob24} in blue and our BCGs plotted as red points. We see that contrarily to the $\mathrm{M}_\mathrm{BH} - \sigma$ relation, our core-elliptical do follow this scaling relation. Instead, in the bottom panels of Fig.~\ref{Fig.mbh_mgal} we do the same for the $\mathrm{r}_\mathrm{c} - \mathrm{M}_\ast$ correlation of \citet{Rob24}. We see that
the most BCGs agree well with this relation - this is expected given the tightness of the $\mathrm{M}_\mathrm{BH} - \mathrm{r}_\mathrm{c}$ relation - with few outliers in the right panels for which $\mathrm{M}_\ast$ is too high. This might be due to contamination from ICL \citep{Matthias21}. In fact, the effect is removed by evaluating $\mathrm{M}_\ast$ applying the prescriptions of \citet{Matthias21} (see e.g. top and right panels of their Fig. 1) rather than integrating the whole 2D profile. In particular, A399 has a very large M/L value, whereas A2319 has one of the smallest core radii of the sample. \\
We stress here the importance of these correlations: $\mathrm{r}_\mathrm{c} - \mathrm{M}_\ast$ allows us to estimate the stellar mass of a galaxy only relying on photometric analysis, whereas $\mathrm{M}_\mathrm{BH} - \mathrm{M}_\ast$ can be used to hunt UMBHs in core-less galaxies such as A2107 or A1201 (see also Sec.~\ref{Sec.core_less} below). However, while the stellar mass still correlates with black hole mass at the high-mass end, the corresponding relation exhibits significantly larger scatter than the one based on core size. These conclusions are in good agreement with the results of \citet{Dullo19}, who finds similar trend using galaxy magnitudes instead of stellar masses for a sample of 41 Core-S{\'e}rsic galaxies, albeit with larger scatter.

\subsection{$r_{SOI} - r_c$} \label{Ssec.soi_core}
\begin{figure}
    \centering
    \includegraphics[width=1.0\linewidth]{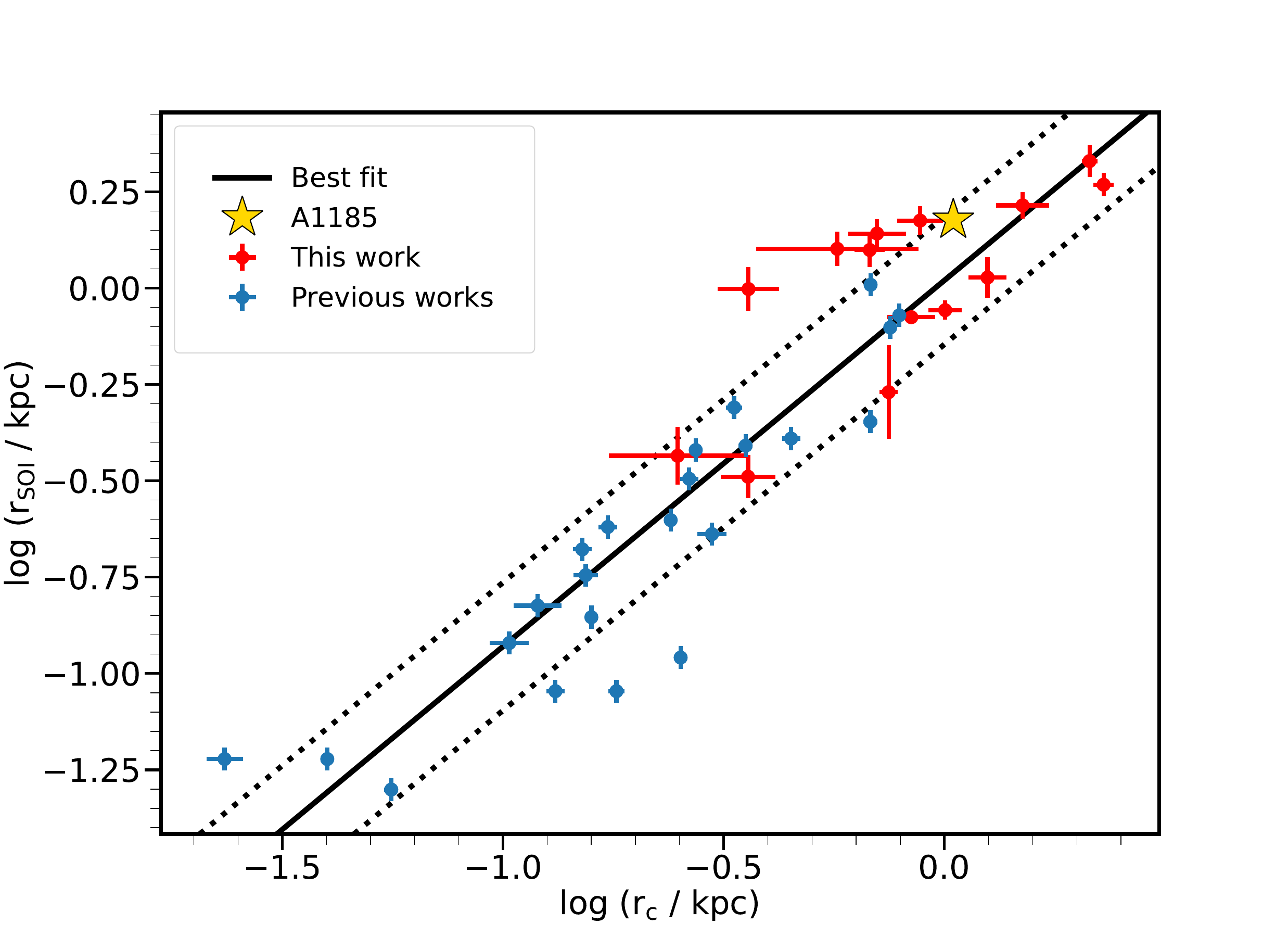}
    \caption{Same as Fig.~\ref{Fig.BH_core} but the correlation between black hole sphere of influence $\mathrm{r}_\mathrm{SOI}$ and core size $\mathrm{r}_\mathrm{c}$ is shown.}
    \label{Fig.rcore_rSOI}
\end{figure}

While the correlation between core size and black-hole mass discussed in the previous Sec.~\ref{Ssec.bh_core} seems natural -- bigger black-holes carve out bigger cores -- it is not a direct consequence of the black-hole binary model for core scouring. In fact, a purely dynamical argument supporting the formation of cores by SMBHs is provided by the very tight correlation between the size of the sphere of influence r$_\mathrm{SOI}$, defined as M$_*(< \mathrm{r}_\mathrm{SOI}) = \mbhm$, and the size of the core \citep{Jens16}. In Fig.~\ref{Fig.rcore_rSOI} we demonstrate that this correlation also extends to the regime of UMBHs. The best-fit relation is
\begin{equation}
    \log \frac{r_\mathrm{SOI}}{\mathrm{kpc}} = (0.960 \pm 0.060) \log \frac{\mathrm{r}_\mathrm{c}}{\mathrm{kpc}} + (0.028 \pm 0.039)
    \label{eq.rbsoi}
\end{equation}
with an intrinsic scatter of $\varepsilon = 0.161\,\pm 0.026$.
This is in very good agreement with the relation in \citet{Jens16}.
The fact that $\mathrm{r}_\mathrm{SOI}$ and $\mathrm{r}_\mathrm{c}$ are essentially identical provides the most direct link between the gravitational influence of the SMBHs and cores.

\subsection{$\rho_{core} - r_c$} \label{Ssec.rho_core}
Physically, the core size\footnote{Note that we do not distinguish between intrinsic and projected core size here, since both are very similar in massive elliptical galaxies.} cannot be exclusively determined by \mbh. While the black-hole mass (in an equal mass merger) roughly defines the available gravitational energy of the binary, the size of the depleted core that can be produced as a result of the release of this energy will also depend on the binding energy of the progenitor centers, i.e. on the core density $\rho_c$. The tightness of the \mbh-r$_c$ relation therefore requires an additional physical ingredient beyond SMBH scouring.

\subsubsection{Core Homology}
Consider a broken power-law density profile for galaxy $i$, with core radius $r_{c,i}$, inner slope $\gamma_i$ inside the core, and outer slope $\alpha_i$ outside it:
\begin{equation}
\rho_i(r) =
\begin{cases}
\rho_{c,i}\left(\dfrac{r}{r_{c,i}}\right)^{-\gamma_i}, & r \le r_{c,i} \\[4pt]
A_i\, r^{-\alpha_i}, & r > r_{c,i}.
\end{cases}
\label{eq:rhodef}
\end{equation}

The central core parameters are connected via the mass enclosed within the core radius, which reads
\begin{equation}
M_{*,i}(r\le r_{c,i}) = \frac{4\pi\,\rho_{c,i}}{3-\gamma_i}\,r_{c,i}^{3}.
\label{eq:enclmass}
\end{equation}

Since $r_{\rm SOI}=r_c$ and $M_*(r\le r_{\rm SOI})\equiv \text{M}_\text{BH}$ by definition of
the sphere of influence, the left-hand side of Eq.~\ref{eq:enclmass} equals
$\text{M}_\text{BH}$. Combined with the empirical relation
$\text{M}_\text{BH} = {\rm M}_{\text{BH},\rm ref}\,r_c^{0.916}$, this gives
\begin{equation}
\rho_{c,i} = \frac{(3-\gamma_i)\,{\rm M}_{\text{BH},\rm ref}}{4\pi}\; r_{c,i}^{\,0.916-3}.
\label{eq:rhob}
\end{equation}

The exponent in Eq.~\ref{eq:rhob}, $3-0.916=2.084$, is fixed entirely by the slope
of the $\text{M}_\text{BH}$--$r_c$ relation, because $\gamma_i$ enters the enclosed-mass (eq.~\ref{eq:enclmass}) only through the prefactor $(3-\gamma_i)$, but not through the exponent on $r_c$. \\
In Fig.~\ref{Fig.rho_core} we show the core densities $\rho_\mathrm{core}$ and core sizes $\mathrm{r}_\mathrm{c}$ of our sample galaxies explicitly. As expected, there is a strong correlation
\begin{equation}
    \log \frac{\rho_\mathrm{core}}{M_\odot \mathrm{kpc}^{-3}} = (-2.08\,\pm\,0.11) \log \frac{\mathrm{r}_\mathrm{c}}{\mathrm{kpc}} + (9.313\,\pm\,0.068).
    \label{eq.rhobrb}
\end{equation}
with an intrinsic scatter of $\varepsilon = 0.272\,\pm 0.043$. Eq.~\ref{eq.rhobrb} is the 3d version of the observed tight scaling between the core surface brightness and core size \citep{Faber97}. \\
 Our fitted $\rho_c$--$r_c$ slope of $2.08$ is consistent with
this predicted value of $2.084$. This is not surprising: the predicted slope is
insensitive to how $\gamma_i$ varies across the sample, and holds as an algebraic
identity for any $\gamma_i$ - the inner slope affects only the normalization.
Using the fitted amplitudes, $\text{M}_\text{BH} = 10^{10.087}\,r_c^{0.916}$ and
$\rho_c = 10^{9.313}\,r_c^{-2.08}$, we can solve for the mean inner slope directly:
\begin{equation}
\langle\gamma_i\rangle = 3 - 4\pi\,\frac{10^{9.313}}{10^{10.087}} = 0.885,
\label{eq:alpha}
\end{equation}
a plausible value for the average 3D core density slope.

\subsubsection{Progenitor Homology}
\begin{figure*}
    \centering
    \subfloat[\label{Fig.rho_core}]{\includegraphics[width=0.4\linewidth]{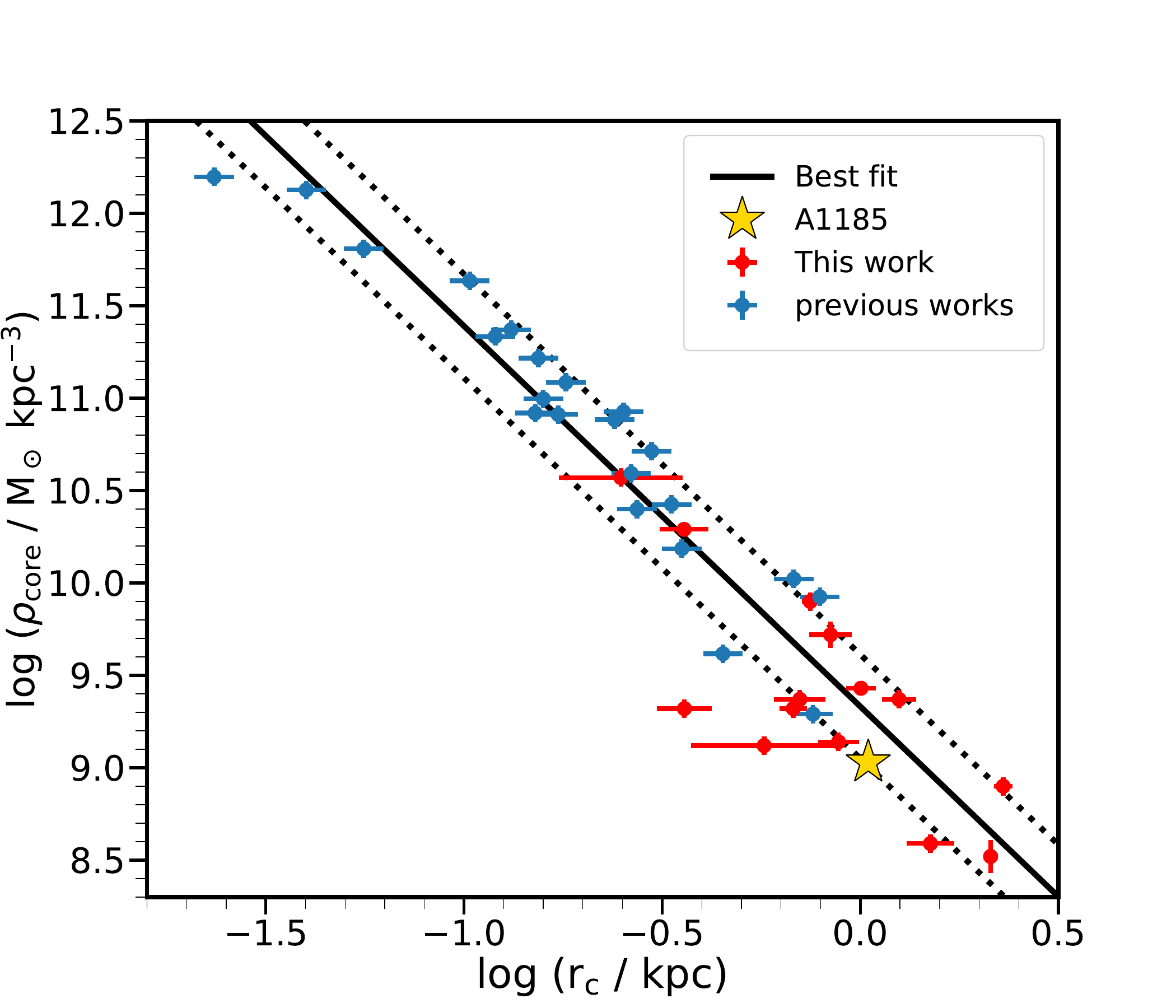}}
    \subfloat[\label{Fig.plaw_dens}]{\includegraphics[width=0.4\linewidth]{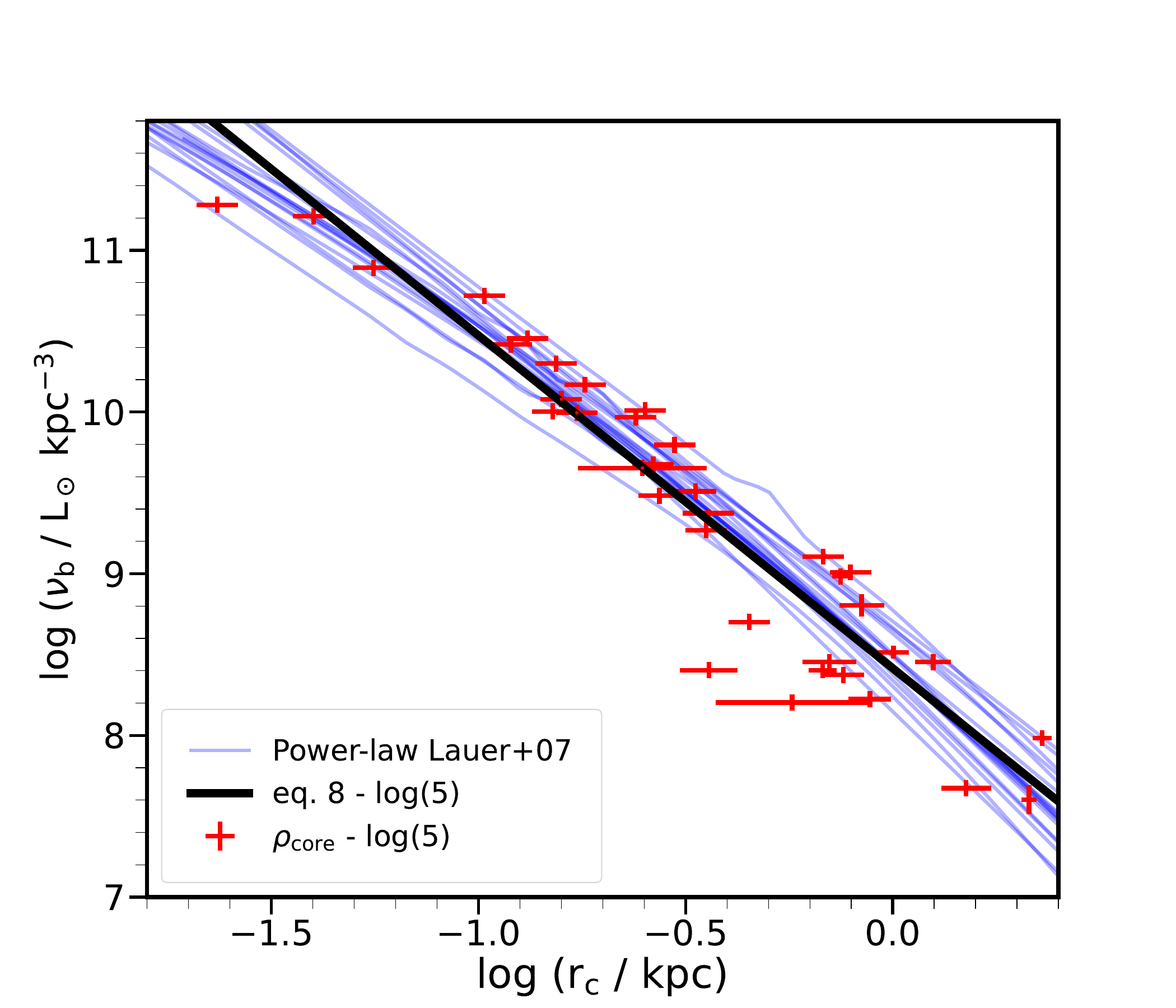}}
    \caption {\textit{Left}: Same as Figs.~\ref{Fig.BH_core} and \ref{Fig.rcore_rSOI} but the correlation between core density $\rho_\mathrm{core}$ and core size $\mathrm{r}_\mathrm{c}$ is shown. \textit{Right:} Spherical deprojections $\nu_\text{b}$ of the 18 brightest power-law ellipticals from \citet[blue lines]{Lauer07} with the data points from the left panels shown as green points as well as eq.~\ref{eq.rhobrb}. A common mass-to-light ratio of 5 was applied. The $\rho_\mathrm{core}$-$\mathrm{r}_\mathrm{c}$ relation is strikingly similar to the density profiles of power-law galaxies in both slope and normalisation.}
    \label{Fig.rho_core_twopanels}
\end{figure*}

The same empirical $\rho_c$--$r_c$ relation is far more consequential for the interpretation of the outer, potentially pre-scouring parameters than for the inner slope, even though the two branches meet at exactly the same point, $r_c$: because $r_{c,i}$ is
set by the scouring dynamics of each galaxy's SMBH binary -- not by anything intrinsic to its outer density distribution -- there is no obvious reason why $\rho_i(r_{c,i})$ should correlate with $r_{c,i}$ across galaxies at all. Mathematically, unlike for the inner slope, there is no equation analogous to eq.~\ref{eq:enclmass} tying $A_i$ and $\alpha_i$ to $r_c$; so, if we evaluate the
outer branch of eq.~\ref{eq:rhodef} at $r=r_{c,i}$ and insert the empirical relation
$\rho_{c,i}=\rho_{c,\rm ref}\,r_{c,i}^{-2.084}$, what remains is
\begin{equation}
A_i = \rho_{c,\rm ref}\; r_{c,i}^{\,\alpha_i - 2.084}.
\label{eq:Ai}
\end{equation}
This in general requires the outer normalization $A_i$ to depend on $r_{c,i}$
through whatever $\alpha_i$ happens to be. A particularly simple case arises, however, if the outer slope is itself common to all galaxies: setting $\alpha_i\approx2.084$ for all $i$ -- equal to the slope of the $\rho_c$--$r_c$ relation itself -- is the unique choice for which $A_i$ automatically becomes the same for all galaxies as well. We call this the \textit{homology solution}. Real coreless galaxies lie close to this homology solution\footnote{Equivalently, since $r_{\rm SOI}\approx r_c$ and $\text{M}_\text{BH}\propto r_c$, the same relations say that the enclosed mass grows linearly with radius across the sample at the core radius. The homology solution $\rho(r)\propto r^{-2}$ is the profile for which this holds without any additional tuning between galaxies.}, as we demonstrate in Fig.~\ref{Fig.plaw_dens} by showing the spherical deprojections of the 18 brightest power-law ellipticals from \citet{Lauer07}. Furthermore, since dissipationless merger remnants predominantly inherit the steepest density slope of their progenitors \citep{Fulton01,Dehnen05}, the striking similarity between the inter-galactic $\rho_\mathrm{core}$-$\mathrm{r}_\mathrm{c}$ scaling law and the intra-galactic density slopes of their likely progenitors (Fig.~\ref{Fig.plaw_dens}) is not a coincidence but suggests a natural homology among the latter, in agreement with previous findings on similar inner light profiles (see e.g. Fig. 12 of \citealt{Dullo12}). \\
Eq.~\ref{eq:Ai} does not by itself rule out a population of individually
non-homologous profiles, with $A_i$ and $\alpha_i$ varying together in just the way needed to reproduce the same relation collectively. Simulations will have to show whether such a construction could be matched to the observed relations, or whether progenitor homology is a natural requirement for the observed core correlations to hold over four orders of magnitude in density.


\subsection{M$_{BH}$-$M_{def}$} \label{Ssec.mbh_mdef}
\begin{figure*}
\centering
\subfloat[\label{Fig.mbh_mdef_scatter}]{\includegraphics[width=0.4\linewidth]{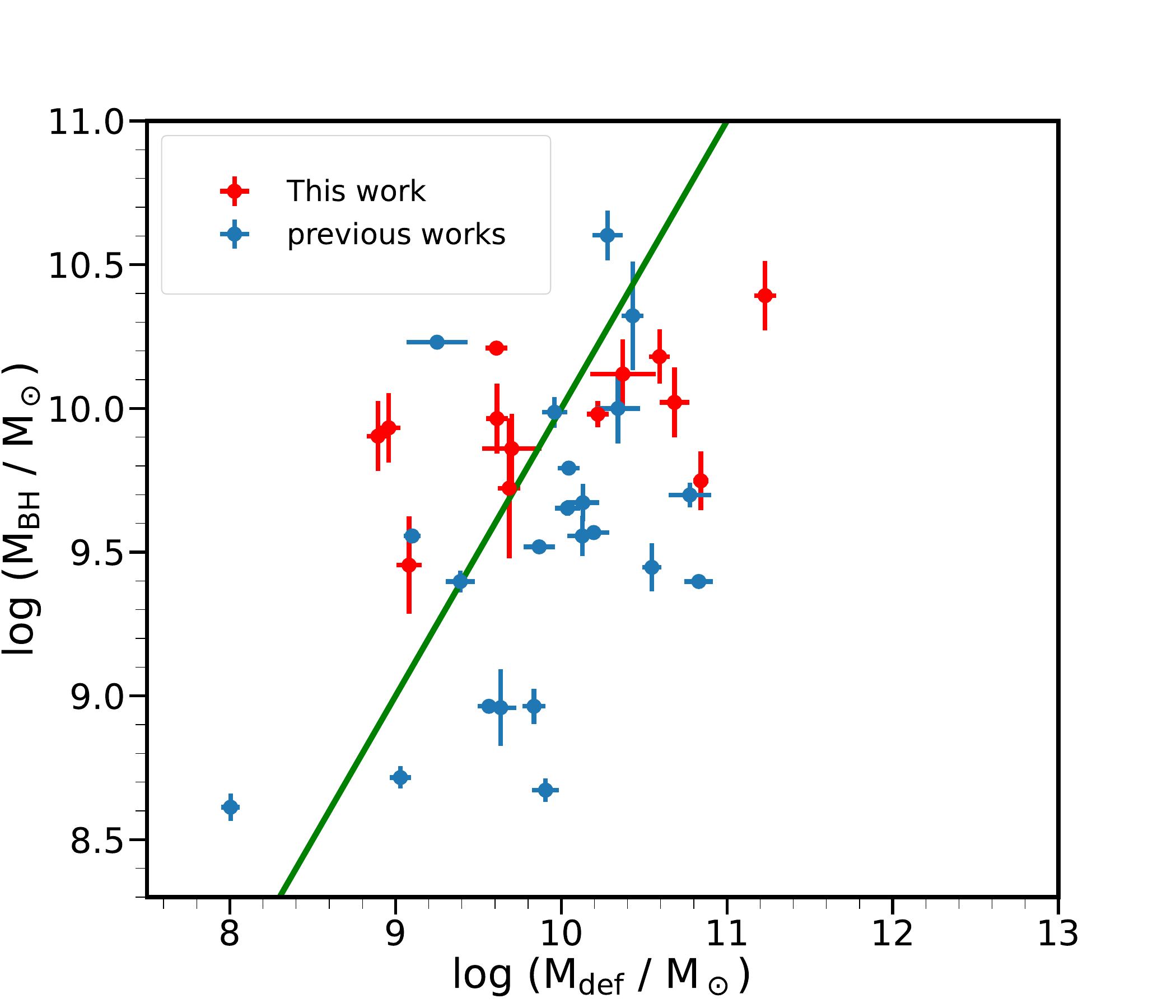}}
\subfloat[\label{Fig.mbh_mdef_hist}]{\includegraphics[width=0.4\linewidth]{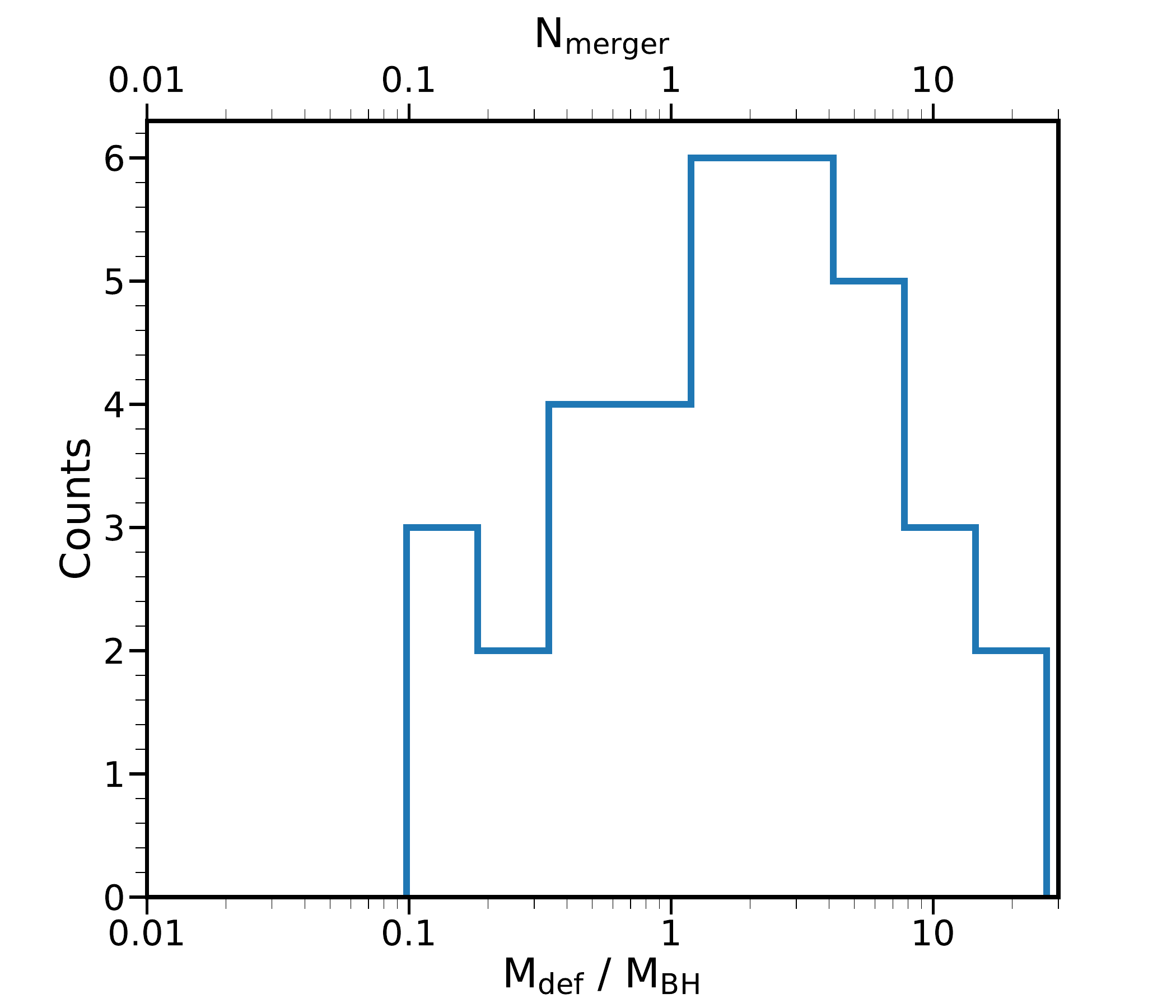}}

    \caption{\textit{Left}: Plot of the mass deficit M$_\mathrm{def}$ inside the core against \mbh. The green line indicates the 1:1 relation. No strong correlation is found. \textit{Right}: M$_\mathrm{def}$ values scaled to the respective \mbh\,estimates. The top axis shows the smallest number of mergers needed to generate M$_\mathrm{def}$ according to \citet{Merritt06} assuming an efficiency of 1$\mbhm$ per merging event. The blue points are from \citet{Rusli13cores}.}
    \label{Fig.mbh_mdef}
\end{figure*}

A final remark concerns the possible correlation between \mbh\,and the mass deficit M$_\mathrm{def}$ inside the core. This measures how much mass has been expelled from the core during its formation and is computed by taking a Sersic profile which reproduces the SB profile outside the core and integrate the luminosity difference L$_\mathrm{def}$ between this Sersic function and the best-fit Core-Sersic up to infinity using the circularized radius (see \citealt{Graham04, Rusli13cores}). The missing light in the centre can then be turned into M$_\mathrm{def}$ using the dynamical mass-to-light ratio\footnote{The mass deficits are calculated for the stellar component only. Since our dynamical models include a separate dark matter potential ($\rho_{\rm DM}$), the scaling factor $\Gamma$ isolates the stellar mass normalization. We therefore do not assume that dark matter is depleted in proportion to the stars during the processes responsible for the observed stellar cores.}. We note that the use of Core-S{\'e}rsic models generally yields mass deficits significantly smaller than earlier estimates based on alternative approaches \citep{Graham04,Dullo12}. \\
\citet{Rusli13cores} computed the mass deficit for a sample of 23 core-galaxies with dynamical \mbh\,estimates (see their Table 4), finding that most galaxies show deficits below 10\mbh. In Fig.~\ref{Fig.mbh_mdef_scatter} we plot the deficits from our sample together with those from \citet{Rusli13cores} and NGC1272 \citep{Rob24}. The correlation between \mbh\,and M$_\mathrm{def}$ is not particularly strong: the Spearman's correlation coefficient is only 0.36, possibly reflecting the limited dynamic range focused on UMBHs. While previous studies \citep{Graham04, Ferrarese06, Kormendy09, Dullo14} over broader mass ranges generally find a stronger correlation, \citet{Rusli13cores} also report only a modest correlation. Fig.~\ref{Fig.mbh_mdef_hist} shows the histogram of the mass deficits scaled to the respective \mbh\,value. Simulations show \citep{Gualandris08} that a single dry merger can produce a mass deficit of up to $5\mbhm$, although this extreme value is rare and corresponds to cases with gravitational-wave recoil. In more typical cases, however, the deficit per major merger is $\sim 0.5$–$1\mbhm$ \citep{Merritt06}, implying that the largest observed deficits may require a substantially larger number of mergers, consistent with the hierarchical assembly expected for BCGs. At the same time, the majority of systems in our sample remain compatible with only a few major dry mergers, in agreement with \citet{Rusli13cores} and in good agreement with the findings of \citet{Dullo19} using the $\mbhm - \mathrm{r}_\mathrm{c}$ relation.



\subsection{Can UMBHs in core-less galaxies exist?} \label{Sec.core_less}

\citet{Nightingale23} report a 30-billion solar mass BH in a core-less galaxy. In this work, we found a second example: A2107 does not show any resolved core \citep{dN26}, has the \textit{second most massive UMBH of the whole sample} and the largest $\sigma$. If the presence of UMBHs does not necessarily require a core elliptical, then the demography and the evolution of high-mass end of the BH mass function cannot be determined from a census of core galaxies but requires dedicated surveys of respective core-less galaxies, with important consequences in the context of  Pulsar Time Array (PTA) data. In fact, a gravitational-wave background is expected from the distribution of merging SMBH binaries. In particular, the properties of the mergers - mass of the progenitors, spin - shape the spectrum and the amplitude of the background \citep{Liepold24}. Such candidates can be identified using the \mbh-M$_\ast$ relation, which - in contrast with the \mbh-$\sigma$ - works well at the high-mass end (see Sec.~\ref{Ssec.mgal}). The remaining question is how these systems can exist. A simple, possible explanation would be the following: while the widely accepted evolution scenario leading to the formation of massive ETGs suggests that these accrete mass through dry mergers in the latest phases \citep{Bender92, Kormendy96}, a small number of galaxies may well have experienced late wet mergers as well or growth from cooling-flow gas, resulting in cuspy profiles \textit{and} UMBHs. We note that alternative mechanisms, such as the \textit{galforming} scenario introduced by \citet{Graham23}, have been proposed to explain UMBHs in galaxies without obvious depleted cores.

\begin{table*}
\centering
\makebox[\textwidth][c]{%
\resizebox{\textwidth}{!}{%
\begin{tabular}{ccccccccccccc}
\hline
Galaxy & ID & Distance (Mpc) & M$_\mathrm{BH}$ (10$^9$ M$_\odot$) & $\mathrm{r}_{\mathrm{c}}$ (kpc) & $\mathrm{r}_\mathrm{SOI}$ (kpc) & $\sigma$ (km s$^{-1}$) & $\rho_\mathrm{core}$ (10$^9$ M$_\odot$ kpc$^{-3}$) & M$_\mathrm{gal,2D}$ (10$^{11}$ M$_\odot$) & M$_\mathrm{gal,Sersic}$ (10$^{11}$ M$_\odot$) & M$_\mathrm{def}$ (10$^{9}$ M$_\odot$) \\
\hline
A160 & MCG+02-04-010 & 179.24 & $13.2 \pm 3.7$ & 0.88 $\pm$ 0.10 & 1.50 $\pm$ 0.13 & 302.3 $\pm$ 7.3 & 1.380 $\pm$ 0.069 & 25.4 $\pm$ 2.3 & 8.9 $\pm$ 1.2 & 23.5 $\pm$ 10.7\\
A240 & UGC1191 & 250.00 & $9.2 \pm 2.6$ & 0.678 $\pm$ 0.053 & 1.26 $\pm$ 0.13 & 258.2 $\pm$ 4.3 & 2.09 $\pm$ 0.10 & 10.9 $\pm$ 1.1 & 4.9 $\pm$ 1.0 & 4.09 $\pm$ 0.61 \\
A292 & UGC1518 & 257.83 & $13.8 \pm 3.9 $ & 1.25 $\pm$ 0.12$^{\ast\ast}$ & 1.06 $\pm$ 0.13 & 311.0 $\pm$ 7.3 & 2.34 $\pm$ 0.12 & 33.3 $\pm$ 2.3 & 25.2 $\pm$ 3.9 & - \\
A399 & UGC2438 & 281.96 & $7.2 \pm 2.0$ & 1.50 $\pm$ 0.21 & 1.64 $\pm$ 0.13 & 289 $\pm$ 10 & 0.389 $\pm$ 0.019 & 100 $\pm$ 11 & 32.6 $\pm$ 3.6 &  5.03 $\pm$ 0.74 \\
A592  & 2MASXJ0742 & 261.54 & $8.6 \pm 2.4$  & 0.57 $\pm$ 0.24 & 1.26 $\pm$ 0.13 & 278.3 $\pm$ 5.1 & 1.318 $\pm$ 0.066 & 55.8 $\pm$ 3.7 & 26.2 $\pm$ 6.1 &  0.91 $\pm$ 0.14\\
A634   & UGC4289 & 113.03 & $5.6 \pm 1.3$  & 0.249 $\pm$ 0.088 & 0.367 $\pm$ 0.063 & 266.9 $\pm$ 4.1 & 37.2 $\pm$ 1.9 & 8.07 $\pm$ 0.90 & 9.6 $\pm$ 1.5 & 69.6 $\pm$ 7.1\\
A1185  & NGC3550 & 145.00 & $11.2 \pm 3.1$ & 1.05 $\pm$ 0.10$^{\ast\ast}$ & 1.50 $\pm$ 0.13 & 301.9 $\pm$ 6.4 & 1.072 $\pm$ 0.054 & 12.8 $\pm$ 1.1 & 9.9 $\pm$ 1.2 & - \\
A1314  & IC712 & 137.99 & $5.3 \pm 3.0$ & 0.619 $\pm$ 0.035 & 0.54 $\pm$ 0.15 & 277.3 $\pm$ 5.6 & 7.94 $\pm$ 0.40 & 9.75 $\pm$ 0.34 & 3.98 $\pm$ 0.22 & 4.85 $\pm$ 0.76 \\
A1749  & IC4269 & 224.83 & $10.5 \pm 2.9$  & 0.84 $\pm$ 0.10 & 0.8404 $\pm$ 0.0040 & 286.7 $\pm$ 3.7 & 5.25 $\pm$ 0.37 & 14.25 $\pm$ 0.87 & 10.69 $\pm$ 0.97 & 48 $\pm$ 10 \\
A1775  & MCG+05-32-063 & 297.23 & $15.1 \pm 3.3$ & 2.133 $\pm$ 0.088 & 2.14 $\pm$ 0.20 & 336 $\pm$ 11 & 0.331 $\pm$ 0.030 & 76.0 $\pm$ 4.0 & 12.2 $\pm$ 1.6 & 39.2 $\pm$ 5.6\\
A2107  & UGC9958 & 172.23 & $22.4 \pm 3.3$ & - & 1.04 $\pm$ 0.19 & 373.1 $\pm$ 8.1 & 14.45 $\pm$ 0.87 & 55.9 $\pm$ 2.5 & 24.2 $\pm$ 2.0 & -\\
A2147  & UGC10143 & 146.04 & $16.21 \pm 0.90$ & 0.70 $\pm$ 0.11 & 1.38 $\pm$ 0.12 & 266.3 $\pm$ 6.1 & 2.34 $\pm$ 0.12 & 13.9 $\pm$ 2.3 & 3.04 $\pm$ 0.44 & 4.06 $\pm$ 0.70 \\
A2256  & UGC10726 & 237.62 & $24.7 \pm 6.9$ & 2.29 $\pm$ 0.12 & 1.86 $\pm$ 0.13 & 320.3 $\pm$ 8.1 & 0.890 $\pm$ 0.027 & 62.5 $\pm$ 2.2 & 17.1 $\pm$ 1.5 & 170 $\pm$ 26\\
A2319 & MCG+07-40-004 & 220.29 & $8.0 \pm 2.3$ & 0.360 $\pm$ 0.058 & 0.994 $\pm$ 0.13 & 386.0 $\pm$ 9.3 & 2.09 $\pm$ 0.10 & 95.2 $\pm$ 2.6 & 4.38 $\pm$ 0.91 & 0.78 $\pm$ 0.12\\
A2388  & LEDA140981 & 246.28 & $9.6 \pm 1.0$ & 1.00 $\pm$ 0.087 & 0.876 $\pm$ 0.050 & 234.1 $\pm$ 5.1 & 2.69 $\pm$ 0.11 & 48.5 $\pm$ 2.4 & 9.5 $\pm$ 2.0 & 16.61 $\pm$ 2.46\\
A2506 & NGC7432 & 106.43 & $2.8 \pm 1.1$  & 0.359 $\pm$ 0.051 & 0.324 $\pm$ 0.042 & 235.1 $\pm$ 5.6 & 19.50 $\pm$ 0.39 & 9.01 $\pm$ 0.54 & 5.64 $\pm$ 0.85 & 1.21 $\pm$ 0.21\\
\hline
\end{tabular}
}%
}
    \caption{The 16 BCGs considered throughout this work. \textit{Col. 1}: BCG name; \textit{Col. 2}: Common ID to identify the galaxy; \textit{Col. 3}: Distance; \textit{Col. 4}: Black Hole mass; \textit{Col. 5}: Core size. The double $^{\ast\ast}$ denotes estimates obtained using the cusp radius instead of the break radius; \textit{Col. 6}: BH Sphere of influence from integrated mass profiles;  \textit{Col. 7}: Velocity dispersion; \textit{Col. 8}: Core density; \textit{Cols. 9-10}: Galaxy masses from 2D profiles and ICL-corrected Sersic 1D fits, respectively; \textit{Col. 11}: The mass deficit inside the core.}
    \label{Tab.bh_core}

\end{table*}

\section{Conclusions}
We confirmed that the canonical \mbh-$\sigma$ relation is inadequate for predicting black hole masses at the high-mass end, and the effect is particularly severe for BCGs, given their low central velocity dispersion.
In contrast, the core size predicts \mbh\,much more accurately - by about a factor of two in scatter compared to $\sigma$ - and allows for the rapid identification of galaxies potentially hosting massive BHs using photometric data alone, confirming earlier findings \citep{Rusli13cores, Dullo14, Jens16}. We show this for the first time using a sample of triaxial dynamical BH mass measurements, eight of which exceed $10^{10}$ M$_\odot$.
While systems with large cores and small black holes can exist - with two confirmed detections (A1201 and A2107) - the presented scaling relations and the selection of the targets based only on their core-size, suggest that such systems are rare. On the other side, we do find massive or even ultra-massive black-holes in core-less galaxies, suggesting that even the most massive galaxy can undergo merging processes which can re-form a cuspy SB profile. These galaxies can be identified using the  $\mathrm{M}_\mathrm{BH} - \mathrm{M}_\ast$ relation.

We have also discussed the connection between the $\mathrm{M}_\mathrm{BH} - \mathrm{r}_\mathrm{c}$ correlation and the even tighter correlation between $\mathrm{r}_\mathrm{SOI}$ and $\mathrm{r}_\mathrm{c}$. This provides a direct link to the black-hole binary model of core formation. This model can plausibly explain the $\mathrm{M}_\mathrm{BH} - \mathrm{r}_\mathrm{c}$ correlation and we have shown how the homology properties and in particular the central densities of the progenitor galaxies influence the expected slope of the relation.

We plan to extend the study of cores and of the evolution of black holes and host galaxies to higher redshift, up to $z = 1$, with data from the Euclid telescope. Our discussion of the role of progenitor properties could help providing clues how to interpret any potential evolution in the various $\mathrm{M}_\mathrm{BH}$-core correlations.


\section*{Acknowledgements}
Computations were performed on the HPC systems Raven and Cobra at the Max Planck Computing and Data Facility. \\
The LBT is an international collaboration among institutions in the United States, Italy and Germany. LBT Corporation partners are: LBT Beteiligungsgesellschaft, Germany, representing the Max-Planck Society, the Astrophysical Institute Potsdam, and Heidelberg University; The University of Arizona on behalf of the Arizona university system; Istituto Nazionale di Astrofisica, Italy; The Ohio State University, and The Research Corporation, on behalf of The University of Notre Dame, University of Minnesota and University of Virginia. \\
modsCCDRed was developed for the MODS1 and MODS2 instruments at the Large Binocular Telescope Observatory, which were built with major support provided by grants from the U.S. National Science Foundation's Division of Astronomical Sciences
Advanced Technologies and Instrumentation (AST-9987045), the NSF/NOAO TSIP Program, and matching funds provided by the Ohio State University Office of Research and the Ohio Board of Regents. Additional support for modsCCDRed was provided by NSF Grant AST-1108693.
This paper made use of the modsIDL spectral data reduction reduction pipeline developed in part with funds provided by NSF Grant AST-1108693 and a generous gift from OSU Astronomy alumnus David G. Price through the Price Fellowship in Astronomical Instrumentation. \\
This work makes use of the data products from the HST image archive. The HST data presented in this article were obtained from the Mikulski Archive for Space Telescopes (MAST) at the Space Telescope Science Institute. The specific observations analyzed can be accessed \dataset[here]{https://doi.org/10.17909/cydk-x193}.





\bibliography{bibl}{}
\bibliographystyle{aasjournalv7}



\end{document}